# A laboratory work: A teaching robot arm for mechanics and electronic circuits

Omer Sise[*]

Department of Physics, Kocatepe University, Science and Art Faculty, Afyon, 03200, Turkey

[*]e-mail: omersise@aku.edu.tr

**Abstract**

Mechanics and electronic systems can be applied to the physical models to understand the physical phenomena for students in laboratory. In this paper we have developed a robot arm for a laboratory experiment, where students learn how to design a human arm and fingers with basic knowledge of the mechanics and electronics. This experiment culminates in an exhibition tie together aspects of a surprisingly wide range of disciplines and represents an alternative vision of how robot arm design can be used to teach both physics and electric/electronic engineering. A new tool is described that combines the mechanical arrangement with an electronic control circuit and it is shown that this can be readily used as an instructional tool in the physics laboratory to teach the law of mechanics and basic electronics for both teacher and students.

## 1. Introduction

When we say that physics is an exact science, we mean that its laws are expressed in the form of mathematical equations which describe and predict the results of precise quantitative measurements. The advantage in a quantitative physical theory is not only practical one that it gives us the power accurately to predict and to control natural phenomena. From this point of view, we can design several instruments for a given phenomenon, by using the general laws of physics related to the science and technology. In this way, the purpose of physics education is to give students the necessary set of intellectual tools to live fulfilled lives, not to give narrow mathematical expressions.

At its core, physics with engineering is about designing and building solutions to problems. Interestingly, disciplines such as physics and engineering tend to have more constructionist activities than the sciences, where a large part of the laboratory experience typically involves duplication experiments designed by others.

In Turkey, when students are in the second and third year of high school, they learn more about mechanics in physics classes. Although they learn that electronic components are used to control mechanic devices, it is not easy for them to accept without any direct observation. As a result, there are quite a few students who cannot understand the relationship between the design of the mechanic systems and the control units with electronic circuits. Since they do understand the relationship between these two systems, they learn that how a mechanic system with control unit can be designed. However, it is difficult for them and sometimes they learn these systems only by heart. So they could encounter some difficulties when they see control systems in laboratory or television such as a robot arm and controlling a dc motor with mechanic tools.

Although students could gain a greater understanding of electronics if more time were available for this subject, it is difficult to provide extra time under the present curriculum in Turkey. This means that a more effective and efficient tool is needed in this subject, especially for those who have not studied it enough in high school and university.

Based on the concerns described above, we have developed a new teaching robot arm with fingers which consists of two distinct sections. One is a mechanic system to provide moving in three dimensions and holding objects with four dc-motor, and the other is an electronic control system to start the motion of dc-motors. The point of this work is the combination of these two systems, which means that students can feel relation between mechanic and electronic part by changing the position of the joystick, and can see working of dc-motors and movement of the mechanic part on the robot arm.

## 2. Simulation of a human arm with fingers

A great many of the applications of mechanics may be based directly on Newton's laws of motion. One part of the system studied in this work is designed in this way. Simple law of motions and some electronic device are used. The Hook's law, the balance of the lever and conversion of the circular motion to the linear are examples. Figure 1 shows the main pictures associated with this robot arm. The robot arm is capable of moving in three-dimensional area. It is controlled by a joystick within the duration of a few milliseconds. It is called a robot arm because when the joystick moves in any direction, the arm starts to motion and stops at the same position with the joystick.

A typical tool to illustrate the robot arm consists of two distinct section, mechanic and electronic part. In mechanic part, four dc motors use to provide motion of the arm independently from each other, and electronic part will be described in the following section.

The movement of the robot arm can be seen directly to students by using the joystick. Since students can change the resistance of the potentiometer, they can somehow experience the relation between electronic control and mechanic system by changing the potential difference $V_{AB}$ of the Wheatstone bridge and realising it (see section 3).

The simulation designed as a human arm and fingers are effective for visible phenomena in physics, and a variety of simulation of the natural system has been developed for educational purposes. However, a simulation of a natural system with electronic control sometimes is more complicated and students cannot feel anything about it, and it is too optimistic to conclude that students would understand electronic control system without difficulty if only they could observe the mechanic system.

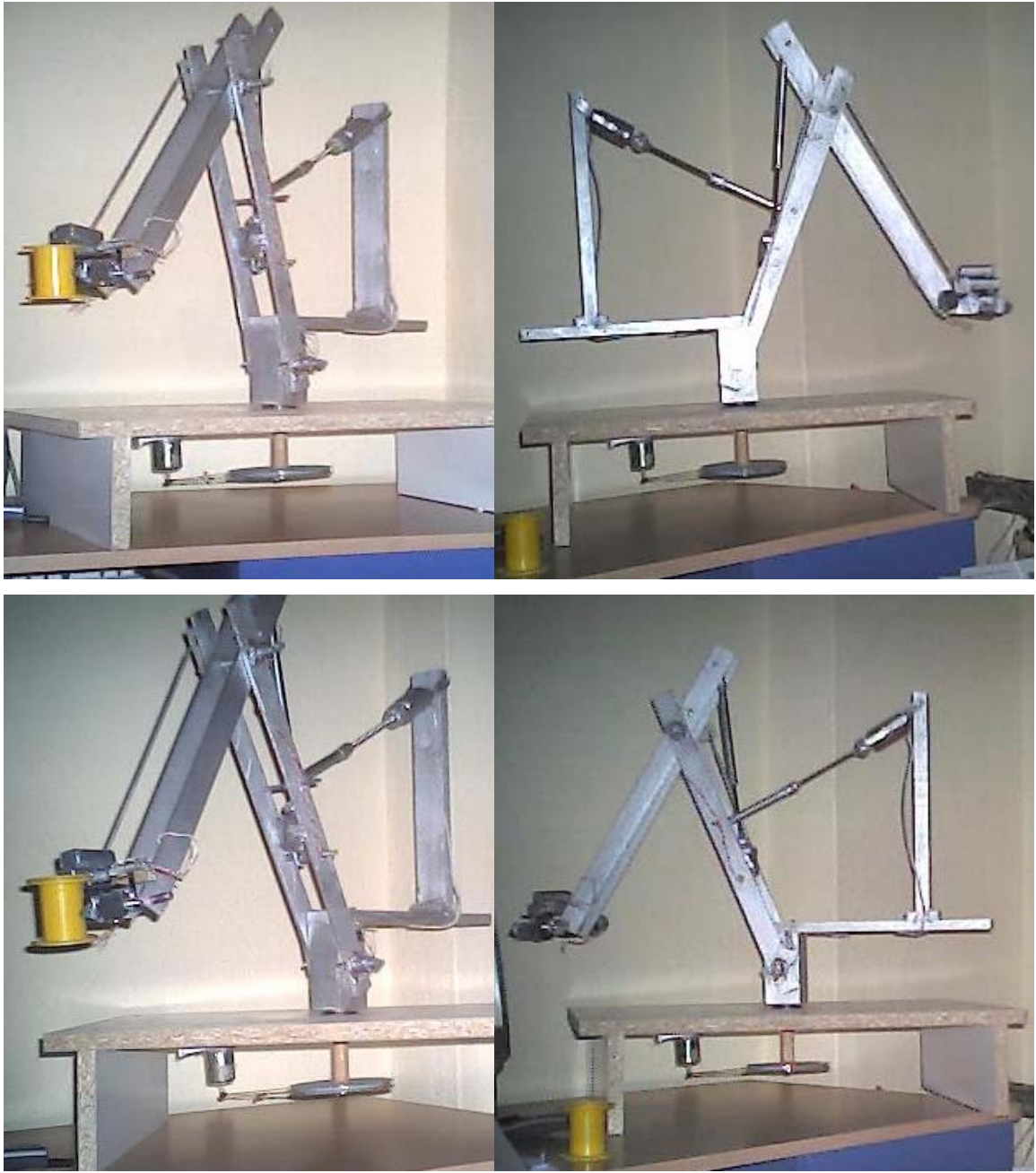

**Figure 1.** Views of the robot arm, and showing connection of screws and motors with and without holding object.

This robot arm shows the balance of a lever and different arrangements are possible. The angle of the robot arm can be mechanically measured at a given length of the screw. Figure 2 shows the angles of mechanic system as a function of the first and second hypotenuse which depend on the length of the screws.

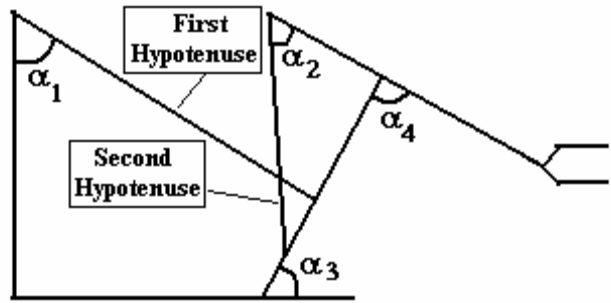

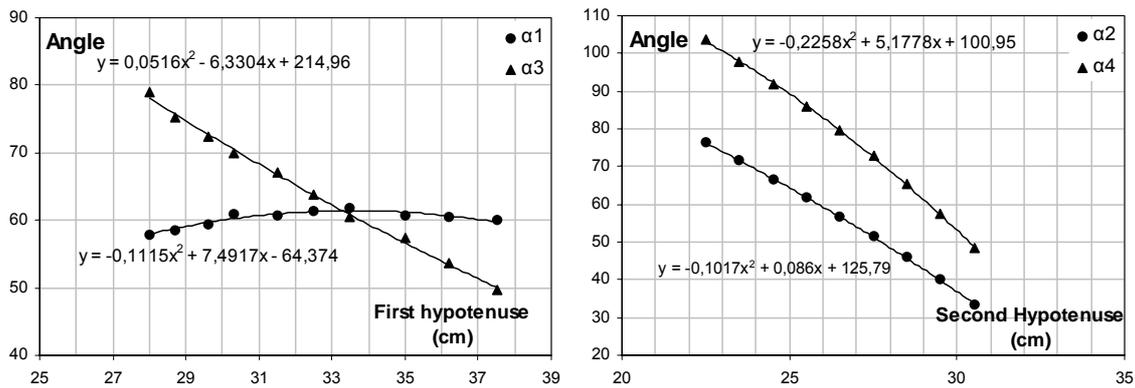

**Figure 2.** Variation of the angle as a function of the (a) first and (b) second hypotenuse which depend on the length of the screws.

## 3. Electronic control

This part sets out to describe an approach to simple circuit involving Wheatstone bridge, transistors, diodes and relays which might be suitable for use. The emphasis is on understanding the basic function of the circuit components; the detailed theory of each component is avoided. The circuit described work very well with n-p-n and p-n-p transistor type BC237 and BC307, respectively. These components can be obtained quite cheaply and is a good transistor for the educational purposes.

We modify the Wheatstone bridge used in this work has two of the potentiometers. It is clear that maximum sensitivity occurs when the equilibrium point is in the middle wire of the potentiometers [1-3]. In figure 3, we start by connecting a 12 V battery and two potentiometers as a Wheatstone bridge, and middle points of the potentiometers, A and B, are connected to the common base and common emitter of two transistors being made positive or negative $V_{AB}$. Two relays are used to switch as an electromagnet. Additionally diodes are connected and parallel to the relays, and they are used to control feedback current.

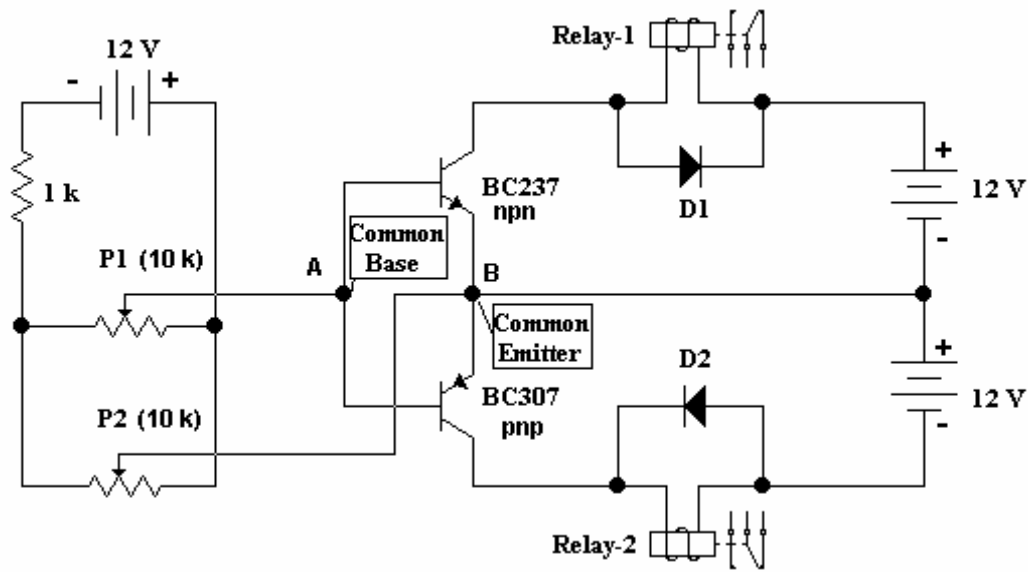

**Figure 3.** Electronic control circuit with Wheatstone bridge.

We now see the effect of varying the position of potentiometers $P_1$ and $P_2$ which controls the potential difference $V_{AB}$ of the Wheatstone bridge. If we start with the same position for potentiometers (50% $P_1$ and 50% $P_2$), $V_{AB}$ will be zero, then no current flows through either relays. As the position of the $P_1$ is changed to the left side (25%), $V_{AB}$ reaches about -2.5 V in pnp region. At this stage a current about 40 mA flows into the emitter of the pnp transistor and the switch of the relay-2 is closed. These cases are valid for the right side (75%) of $P_1$ and then relay-1 works.

Figure 4(a) shows the potential difference of the Wheatstone bridge between A and B as a function of the position of $P_1$ for fixed value of $P_2$. Obtained currents of two regions are also shown in figure 4(b). For other fixed values of $P_2$ (30, 40, 50, 60 and 70%) the current can be determined for two regions as shown in figure 4(c). However, we want the current to vary as much as possible about same value for each npn and pnp region, so we take the position of $P_2$ to be 50% and adjust the potentiometer $P_1$ until the necessary current is obtained.

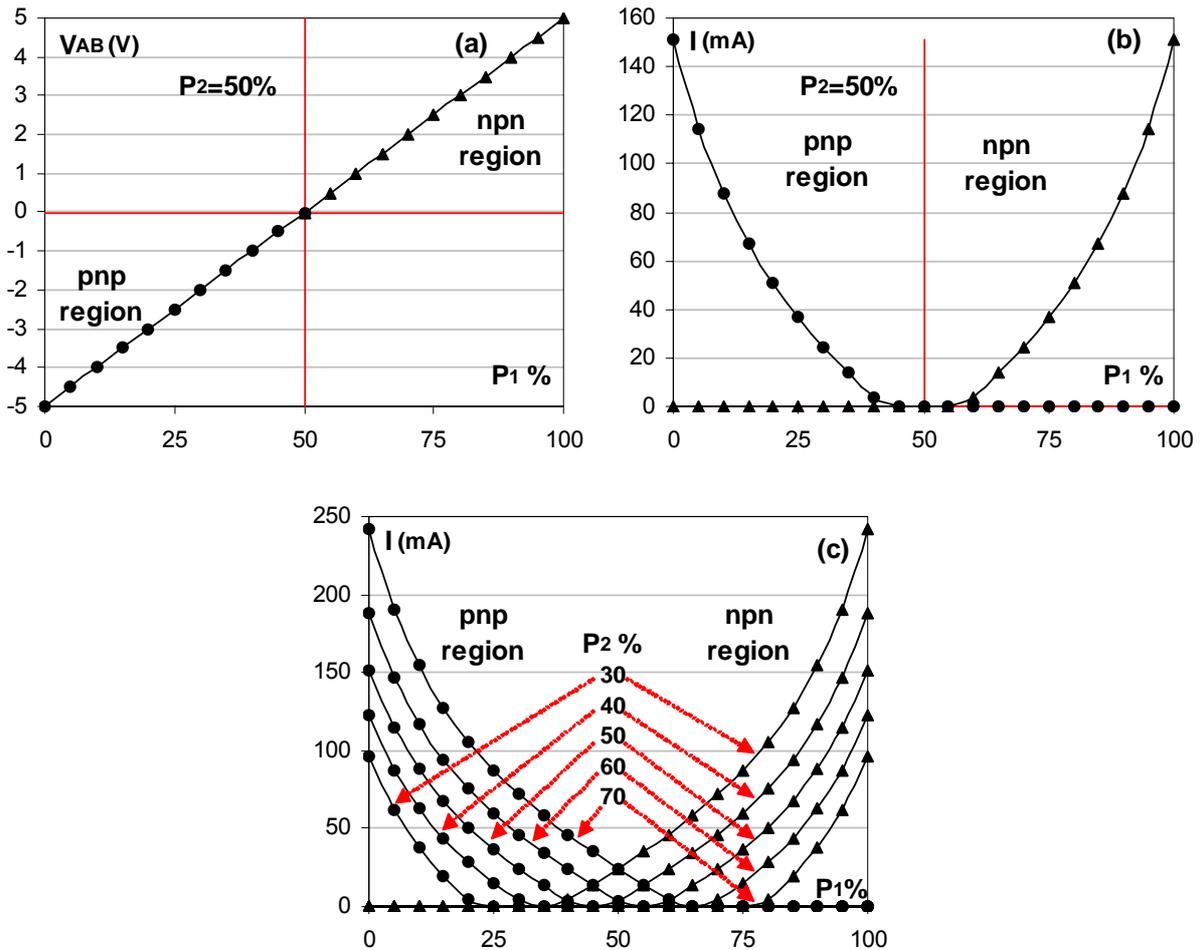

**Figure 4.** (a) Variation of the potential difference between A and B point as a function of $P_1$ for npn and pnp region with $P_2=50\%$. (b) Obtained the current for $P_2=50\%$ and (c) for other values of $P_2$.

The relays which respond to direct current of transistors are necessary to switch outer circuit where a dc-motor connected to the switch of the relays. Two power supplies are used for two directions, that is counter clock-wise (ccw) and clock-wise (cw) direction. For three cases, the working of a dc-motor is shown in figure 5. The potential difference $V_{AB}$ is equal to zero for the same position of the $P_1$ and $P_2$ in figure 5(a) and then relays and dc-motor do not work. When $V_{AB}$ is greater or less than zero, relay-1 or -2 are switched and dc-motor begins to work in ccw or cw direction, respectively (figure 5(b) and (c)).

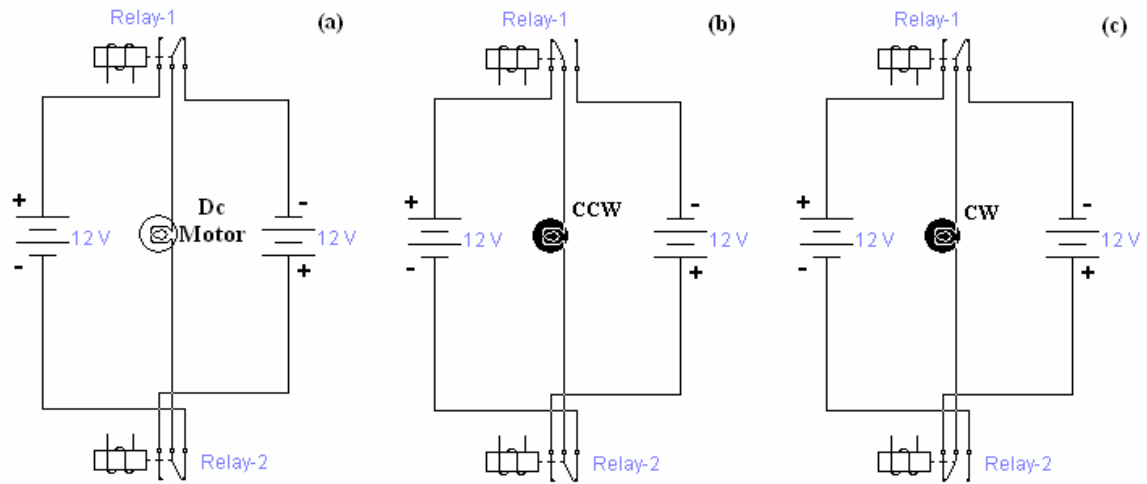

**Figure 5.** The cases of the Wheatstone bridge: (a) $V_A=V_B$, (b) $V_A>V_B$ and (c) $V_A<V_B$

One of the potentiometers is on the fingers and the other is on the joystick where we can control the robot arm at a distance. The external appearance to the electronic control and mechanical arrangement used in this work are presented in figure 6. When the position of the potentiometer on the hand of the joystick is pushed to compress the spring, dc-motor and the potentiometer connected to the screw on the fingers of the robot arm also simultaneously moves according to the potential difference between A and B points. Then when the position of the potentiometer on the robot fingers is equal to the position of the joystick's, the potential difference is zero and the motor does not work.

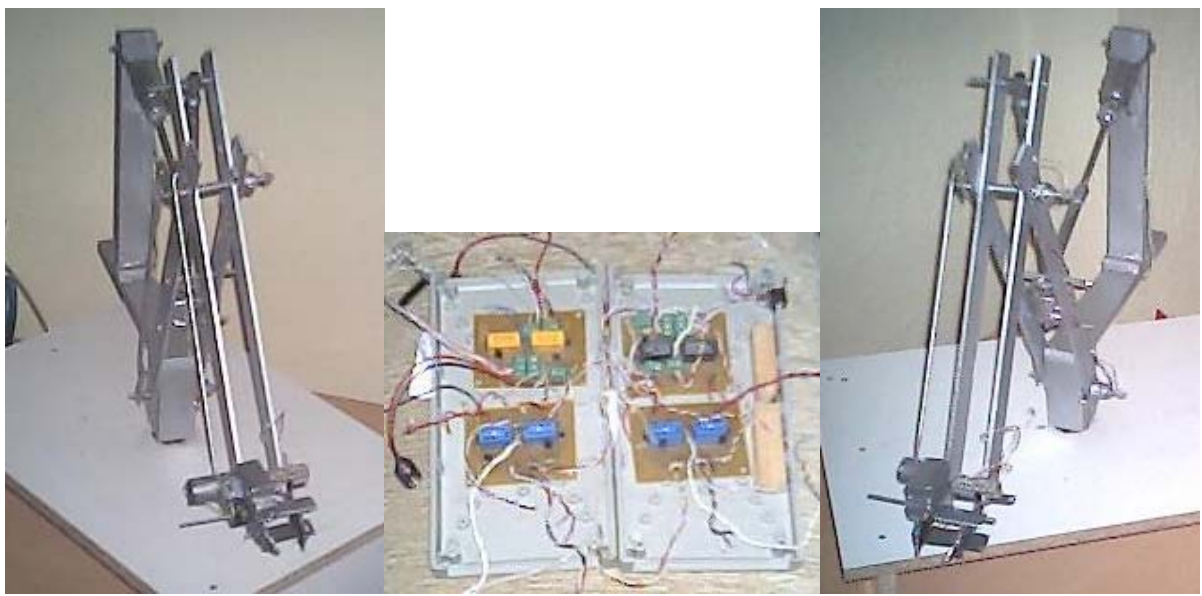

**Figure 6.** Two views of the robot arm with fingers and electronic control circuit for four dc-motors.

We can control the force on robot finger by using the spring. According to Hook's Law, the force exerted by the spring is proportional to the distance the spring is compressed or stretched, $F = kx$, where $k$ is the proportionality constant. Thus the force on the robot arm can be changed by applying different force on the joystick to stretch or compress the spring different distances by our hand. This means that students confirm the potential change physically by hand, visually motion of the holding fingers and simultaneously stopping of dc-motor.

**4. A robot arm as an educational tool**

The robot arm is easy to use and will help physics teachers illustrate concepts such as the law of motion and electronic device in laboratory. It is also most appropriate for high school students studying mechanics and electronic circuits.

Several parameters can be varied. These include the length of the first and second hypotenuse, and the force on the fingers. Best of all, the robot arm can be rotated and viewed from any perspective and students can control the robot arm with different directions. Teachers can use the robot arm in a presentation for laboratory work with students. In one ready-to-use activity, well suited for laboratory or discussion, the teacher can show the robot arm viewed from different perspectives. These presentations provide students with the balance, torque, force on a spring, and the most probable electronic device such as transistor, relay and potentiometers. The student can then learn the controlling of any mechanical system.

**5. Conclusion**

We have developed a new type of a teaching tool, called a robot arm, with which students can relate their physical experience to the conceptual understanding of working of a robot arm by using the law of mechanics and basic electronic knowledge, and showed that how the circuit described works using the minimum number of components and going into the finer points of this design.